\documentclass[aps,
 reprint,
 superscriptaddress,
 amsmath,
 amssymb,
 prl]{revtex4-2}

\usepackage[T1]{fontenc}
\usepackage[utf8]{inputenc} 
\usepackage{graphicx}
\usepackage{dcolumn}
\usepackage{bm}
\usepackage[hidelinks]{hyperref}
\usepackage[mathlines]{lineno}
\usepackage{graphicx}
\usepackage{dcolumn} 
\usepackage{xcolor}
\usepackage[normalem]{ulem}

\begin{document}

\title{Observation of Hexagonal Close-Packed Water Ice at Conditions in Ice Giant Planetary Interiors}

\newcommand{\cea}{CEA DAM DIF, F-91297 Arpajon, France}
\newcommand{\lmce}{Université Paris-Saclay, CEA, Laboratoire Matière en Conditions Extrêmes, F-91680 Bruyères-le-Châtel, France}
\newcommand{\impmc}{Institut de Minéralogie, de Physique des Matériaux et de Cosmochimie (IMPMC), Sorbonne Université,
CNRS UMR 7590, MNHN, 4 place Jussieu, F-75005 Paris, France}
\newcommand{\esrf}{European Synchrotron Radiation Facility, Boîte Postale 220, F-38043 Grenoble, France}

\author{Alexis Forestier}
\email{alexis.forestier@cea.fr}
\affiliation{\cea}
\affiliation{\lmce}

\author{Gunnar Weck}
\affiliation{\cea}
\affiliation{\lmce}

\author{Sandra Ninet}
\affiliation{\impmc}

\author{Gaston Garbarino}
\affiliation{\esrf}

\author{Mohamed Mezouar}
\affiliation{\esrf}

\author{Frédéric Datchi}
\affiliation{\impmc}

\author{Paul Loubeyre}
\affiliation{\cea}
\affiliation{\lmce}

\begin{abstract}
Using synchrotron x-ray diffraction in laser-heated diamond anvil cells, we report the observation of an hexagonal close-packed (hcp) phase of water ice at high pressure and temperature conditions. Above $200$ GPa and $1800$ K, the hcp phase becomes dominant upon entering the superionic regime, as evidenced by anomalous thermal expansion. Observations are consistent with the hcp phase becoming thermodynamically more stable than the face-centered cubic (fcc) phase via a martensitic transition extending across the $130-200$ GPa pressure range, within the superionic regime. Hcp ice is also observed to emerge from stacking disorder developing within the fcc oxygen lattice upon cooling, during its reversion to the bcc phase. The presence of an fcc-hcp martensitic transition in the superionic regime of warm dense ice may have implications for planetary models of Uranus and Neptune.
\end{abstract}

\maketitle

Exploring the phase diagram of water ice under pressure has continuously driven leading-edge experimental developments over the past century \cite{hansen_everlasting_2021}. A large experimental effort is currently devoted to the exploration of the phase diagram of H$_2$O at conditions of planetary interiors, using both static and dynamic experimental techniques \cite{millot_experimental_2018,millot_nanosecond_2019,queyroux_melting_2020,weck_evidence_2022,prakapenka_structure_2021,forestier_x-ray_2025,andriambariarijaona_observation_2025}. 
A symmetric phase, ice X---considered an ionic solid---exists above $60-80$ GPa \cite{polian_new_1984,goncharov_compression_1996,loubeyre_modulated_1999,komatsu_hydrogen_2024}, featuring a body-centered cubic (bcc) oxygen sublattice. At high pressures and elevated temperatures, ice was conjectured to exhibit a superionic state in which oxygen atoms remain on lattice sites while hydrogen nuclei diffuse through the oxygen sublattice \cite{cavazzoni_superionic_1999}.  Remarkably, such a superionic state extends the stability of ice in temperature, making this state directly relevant for ice giant planetary interiors \cite{cavazzoni_superionic_1999}. Theoretical calculations have also predicted ice phases with close-packed oxygen sublattices beyond bcc ice X—either (i) on the cold compression curve in the multi-megabar regime \cite{benoit_new_1996, militzer_new_2010, hermann_high_2012}, or (ii) at high temperature within the superionic state at lower pressure \cite{wilson_superionic_2013, sun_phase_2015, cheng_phase_2021, reinhardt_thermodynamics_2022}. These compact ice phases differ essentially in the stacking sequence of the close-packed oxygen layers, adopting either a face-centered cubic (fcc),  an hexagonal close-packed (hcp) configuration or slightly distorted variants, \emph{e.g.} the \emph{Pbcm} structure \cite{benoit_new_1996}.


The first experimental evidence for the fcc phase, designated as ice XVIII, was obtained via multi-shock compression above 150 GPa and 1500 K  \cite{millot_nanosecond_2019}.  In fact, this phase is the first for which superionic conduction was documented, via optical measurements in laser-driven shock experiments on pre-compressed ice VII \cite{millot_experimental_2018}. The stability of the fcc phase was confirmed and further investigated by synchrotron x-ray diffraction (XRD) in the laser-heated diamond anvil cell (DAC) \cite{weck_evidence_2022,prakapenka_structure_2021}. More recently, fcc ice was also observed by ultrafast x-ray heating of ice within the DAC at the European X-ray Free Electron Laser \cite{husband_phase_2024}. Finally, the implementation of advanced sample geometries in the DAC provided better constraints on the fcc stability field and of the superionic transition, revealed by an anomalous thermal expansion between the insulating and superionic regimes \cite{forestier_x-ray_2025}.
In a recent dynamic experiment, the existence of a fcc-hcp mixed-stacking ice state was inferred from the broadening of one diffraction peak common to both fcc and hcp structures \cite{andriambariarijaona_observation_2025}. 
Here, we report the unambiguous observation of a novel H$_2$O ice phase adopting an hcp oxygen sublattice, based on the identification of five classes of diffraction peaks. In addition we provide evidence that the fcc to hcp transition is of martensitic type and that the hcp phase, like fcc, is a type-II superionic conductor at high temperature.

Sections S1 and S2 of the Supplemental Material (SM) \cite{SuppMat} present a detailed description of the sample preparation and experimental procedures. Briefly, the present experiments used the same sample geometry of an encapsulated ice sample within two cup-shaped boron-doped diamond (BDD) laser absorbers.
As demonstrated in our previous work \cite{forestier_x-ray_2025}, this sample confinement enables efficient indirect laser heating of the ice sample, minimizing temperature gradients thanks to the enclosing geometry and the high thermal conductivity of BDD absorbers \cite{weck_determination_2020}. 
Thermal insulation of the BDD from diamond anvils was ensured by Al$_2$O$_3$ insulating layers. 
X-ray diffraction data were obtained using the sub-micrometer beam of the ID27 beamline of the European Synchrotron Radiation Facility \emph{Extremely Brillant Source} (ESRF-EBS) \cite{mezouar_high_2024}.
The $0.3738$ \AA{} x-ray beam had a spot size of $0.5\times0.8$ µm$^2$ on the sample, ensuring high-quality XRD data despite the small sample size ($\sim 12$ µm diameter at highest pressures above $200$ GPa). 
The beamline uses a two-sided achromatic pyrometric system to determine temperature from the BDD thermal emission spectrum \cite{mezouar_high_2024}. 
Sample pressure was evaluated from the volume of the BDD absorbers using the thermal equation of state of diamond \cite{dewaele_high_2008}.

\begin{figure}[ht]
    \centering
    \includegraphics[width=1\columnwidth]{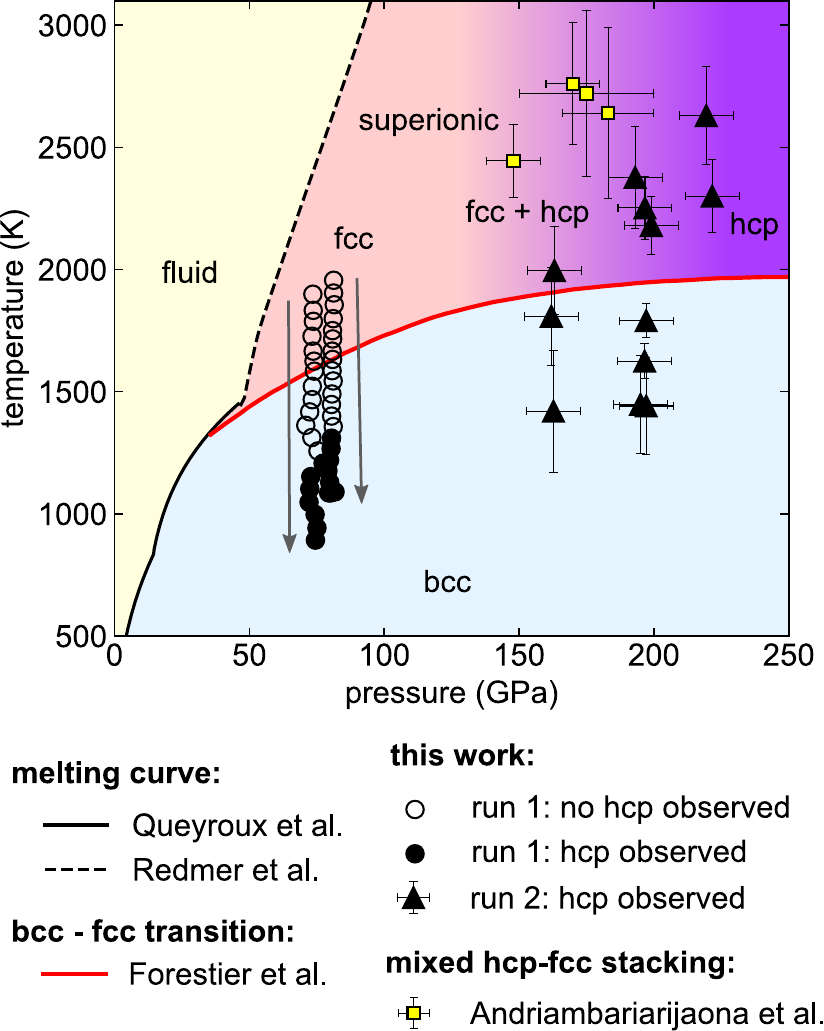}
        \caption{Pressure-temperature diagram showing the thermodynamic conditions of our measurements in runs 1 and 2 (circles and triangles, respectively). Filled (open) symbols indicate the presence (absence) of hcp ice. Note that fcc was observed for all run 1 data points. Gray arrows indicate the two cooling sequences of run 1. For run 1, error bars are omitted for clarity: temperature uncertainties range from $\pm 100$ K at the highest temperatures to $\pm 300$ K at the lowest temperatures, and the pressure uncertainty is $\pm 4$ GPa. Experimental ($P < 50$ GPa) \cite{queyroux_melting_2020} and calculated ($P > 50$ GPa) \cite{redmer_phase_2011} melting curves are shown as solid and dashed black lines, respectively. The bcc-fcc phase boundary from Ref. \onlinecite{forestier_x-ray_2025} is shown as a red line. Observations of fcc + hcp mixed stacking in shock experiments by Andriambariarijaona \emph{et al.} are shown as yellow squares \cite{andriambariarijaona_observation_2025}.}
    \label{fig1}
\end{figure}

\begin{figure}[t]
    \centering
    \includegraphics[width=\columnwidth]{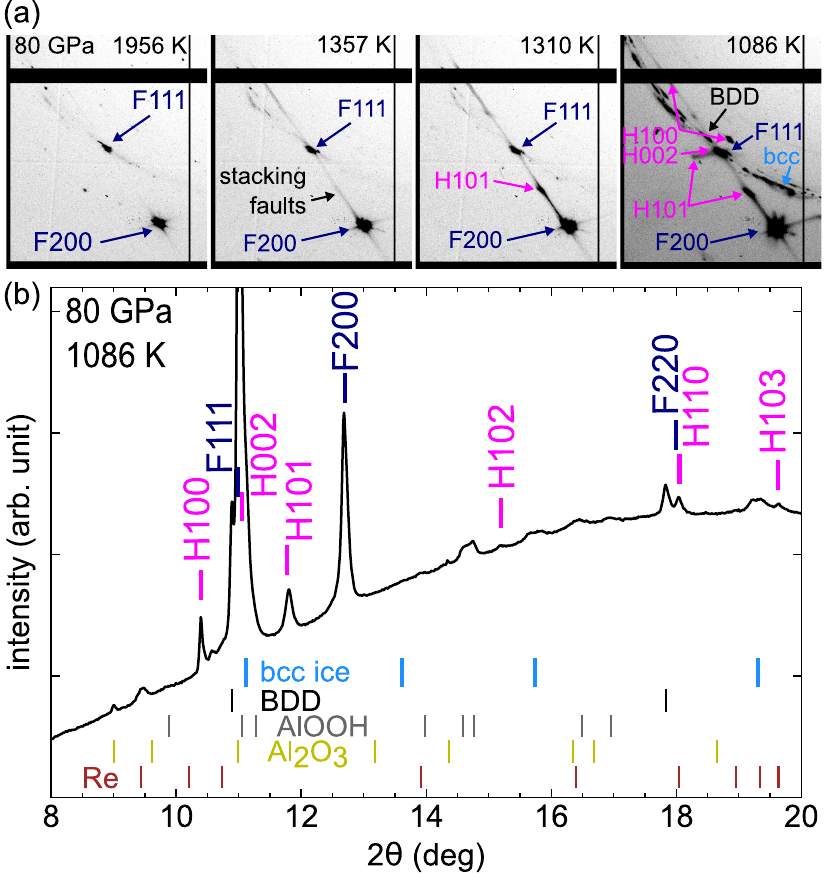}
    \caption{
    (a) Diffraction images from run 1 on hot ice at 80 GPa, recorded while decreasing temperature from 1956 K to 1086 K. The sequence illustrates the evolution of the fcc 111 and 200 reflections (denoted F111 and F200), the onset of x-ray diffuse scattering associated with the development of stacking faults (see text), and the subsequent appearance of hcp Bragg reflections (H100, H002, and H101) at 1310 K and below. At 1086 K, corresponding to the end of the temperature decrease, the final image was collected during a $\pm 10^\circ$ rotation of the DAC.
    (b) Integrated diffraction pattern measured at 80 GPa and 1086 K, corresponding to the last diffraction image in (a). XRD peaks from ice crystals are labeled in pink (hcp) and blue (fcc), while the measured diffraction includes contributions from bcc ice (light blue ticks), the Re gasket (brown ticks), and the BDD heaters (black ticks). Most prominent peaks arising from the Al$_2$O$_3$ insulating layers---corundum, and the $\delta$-AlOOH hydrated phase \cite{duan_phase_2018}---are marked by yellow and gray ticks respectively (see Section 4 of the SM for details on Al$_2$O$_3$ insulating layers evolution at extreme P-T conditions).}
    \label{fig2}
\end{figure}

Two runs were performed: run 1 at $80$ GPa, and run 2 up to $230$ GPa. A pressure–temperature diagram summarizing the thermodynamic conditions of the measurements performed in runs 1 and 2 is presented in Fig. \ref{fig1} (see Section S3 of the SM for a description of the followed thermodynamic paths).
In the first run, we obtained high-quality single crystals of fcc ice by heating the sample up to $1950$ K at $75-80$ GPa. 
Then, diffraction images were collected while cooling the crystal in two successive runs (circles in Fig. \ref{fig1}). As shown in Fig. \ref{fig2}(a), below approximately 1400 K, the fcc diffraction peaks became noticeably deformed and a diffuse signal developed, connecting the fcc 111 and 200 reflections. 
Upon further cooling, well-defined, novel diffraction peaks appeared along the diffuse streaks, which can be unambiguously indexed by a hcp lattice. This evolution indicates a progressive splitting of the fcc 111 and 200 reflections into the hcp 100, 002, and 101 reflections. At 1086 K, transformations remained incomplete, with the coexistence of fcc, hcp and bcc domains. This was consistently observed during the two cooling sequences.
The $d$-spacings corresponding to the fcc 111 and the hcp 002 reflections closely match, indicating that the two structures differ primarily in the stacking sequence of the dense oxygen planes. The fully integrated diffractogram collected at 80 GPa and $1086$ K is shown in Fig. \ref{fig2}(b), displaying the set of Bragg reflections from the fcc lattice ($a_{\mathrm{fcc}} = 3.381$~\AA), together with additional peaks from the hexagonal lattice ($a_{\mathrm{hex}} = 2.383$~\AA, $c_{\mathrm{hex}} = 3.882$~\AA). The latter yields a $c_{\mathrm{hex}}/a_{\mathrm{hex}}$ ratio of 1.629, consistent with the formation of an hcp polytype derived from the parent fcc structure.
The diffuse scattering streaks, labeled \emph{stacking faults} in Fig. \ref{fig2}(a), are characteristic of fcc–hcp stacking faults, as observed in compressed Ar, Kr, and Xe \cite{cynn_martensitic_2001, dewaele_stability_2021, rosa_martensitic_2022, brugman_strength_2025}. 
These features---often called \emph{Bragg rods}---arise from a subset of reciprocal lattice points elongated along the direction normal to planar defects due to the random occurrence of ABAB sequences within the ABCABC stacking of the fcc lattice \cite{petukhov_bragg_2003, edwards_imperfections_1942, wilson_imperfections_1942}.
Here, cooling destabilizes the fcc lattice and promotes the formation of hcp variants. The re-emergence of bcc reflections near $1300$ K upon cooling, concomitant with the emergence of hcp reflections (see Fig. S4 in the SM), suggests that hcp acts as a transient phase in the fcc to bcc transformation.

\begin{figure}[t]
    \centering
    \includegraphics[width=\columnwidth]{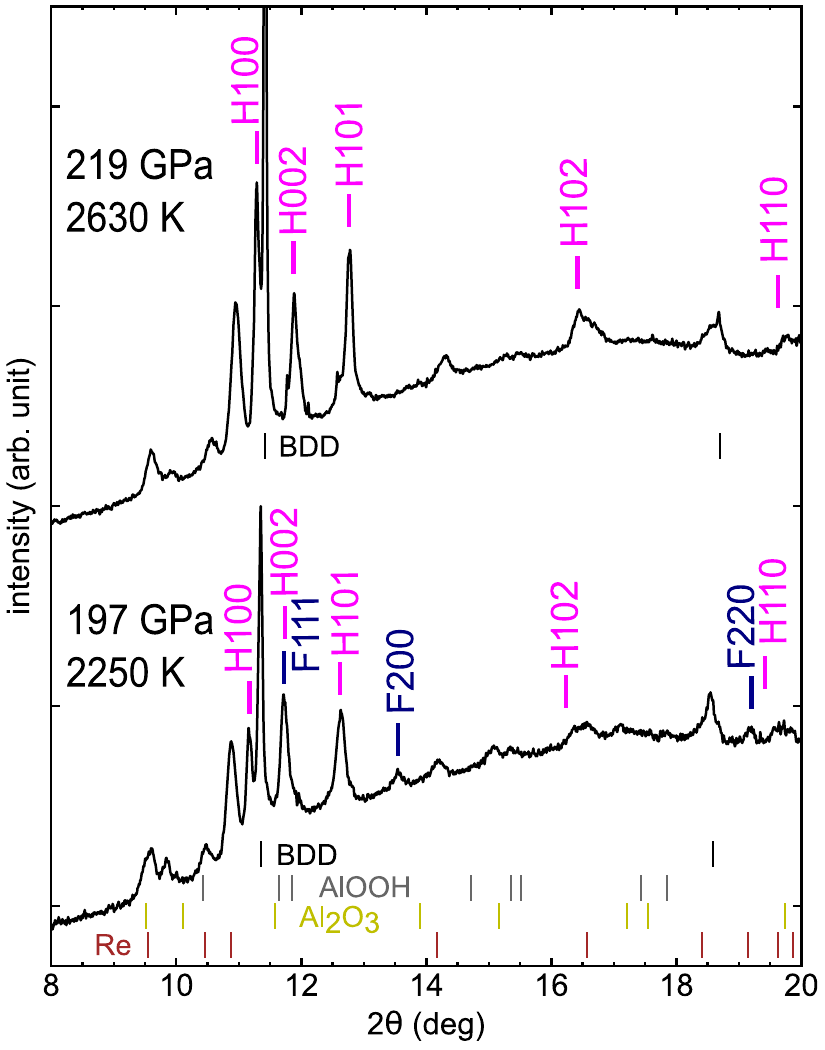}
    \caption{Integrated diffraction patterns collected at 197 GPa, 2250 K (bottom) and 219 GPa, 2630 K (top). As in Fig. \ref{fig2}, diffraction signals from the sample assembly elements are indicated by ticks. 
    In contrast to run 1, here XRD peaks from corundum Al$_2$O$_3$ and $\delta$-AlOOH are faint, as explained in Section S4 of the SM; the most prominent expected peaks are marked by yellow and gray ticks, respectively.} 
    \label{fig3}
\end{figure}

In the second run, the ice sample was initially compressed up to 155 GPa, with thermal annealing cycles at around $1350$ K to relieve non-hydrostatic stresses while maintaining the sample within the stability field of bcc ice X. As expected, no additional diffraction features corresponding to other ice polymorphs were observed at this stage. 
At 155 GPa, a first heating sequence up to $2000$ K led to the appearance of new diffraction peaks, consistent with the coexistence of hcp and fcc. Further compression to $197$ GPa and heating to $2250$ K led to an enhancement of the hcp reflections relative to the fcc ones (Fig. \ref{fig3}, bottom pattern).
As in run 1, the $d$-spacings of the fcc 111 and hcp 002 reflections almost coincide in the XRD patterns at 197 GPa.
Inspection of the corresponding diffraction image (Fig. S6 of the SM), also revealed the diffuse scattering streaks related to the coexistence of hcp and fcc domains in the ice crystals. Finally, during the last heating cycle up to 2630 K and $219$ GPa, the fcc peaks almost vanished and the hcp ice phase clearly dominates (Fig. \ref{fig3}, top pattern). 

Our observations resembles the behavior of compressed noble-gases, in which hcp and fcc polymorphs coexist across a broad pressure range due to stacking disorder associated to martensitic transition mechanisms \cite{cynn_martensitic_2001, dewaele_stability_2021, rosa_martensitic_2022, brugman_strength_2025}.
In fact, recent \emph{ab-initio} simulations showed that the free enthalpies of fcc and hcp (or quasi-hcp) ice oxygen sublattices are similar across a wide thermodynamic P-T range \cite{cheng_phase_2021,reinhardt_thermodynamics_2022,militzer_ab_2018}.

\begin{figure}[t]
    \centering
    \includegraphics[width=.95\columnwidth]{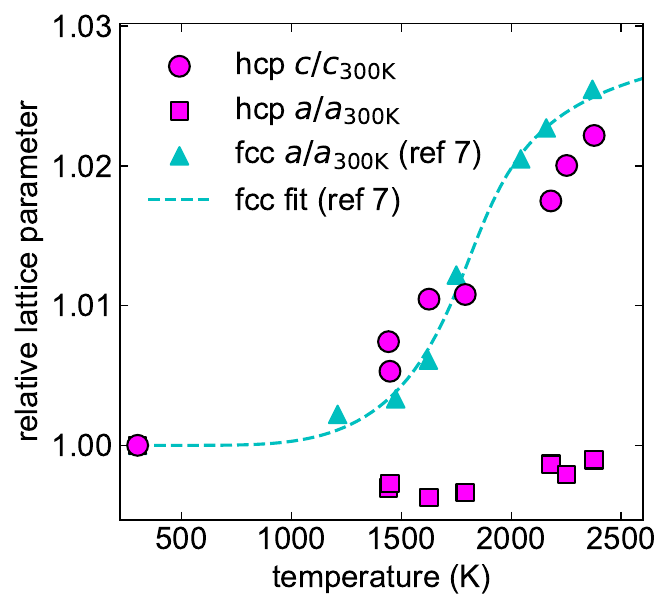}
    \caption{Thermal evolution of the relative lattice parameters $a(T)/a(\mathrm{300~K})$ and $c(T)/c(\mathrm{300~K})$ for the hcp phase near 197 GPa (pink symbols). For comparison, thermal evolution for the fcc ice phase lattice parameter obtained at $191$ GPa from Ref. \onlinecite{forestier_x-ray_2025} are included (blue symbols and dashed line).}
    \label{fig4}
\end{figure}

Fig. \ref{fig4} shows the relative thermal evolution of the lattice parameters for the hcp phase, obtained at different temperatures under laser heating at around $197$ GPa.  Notably, the thermal expansion of the hcp phase is almost entirely driven by the elongation along the c-axis of the unit cell. In fact, the relative variation along this axis mirrors the crossover trend observed for the fcc phase in Ref. \onlinecite{forestier_x-ray_2025}, also included in Fig. \ref{fig4} for comparison. 
The thermal expansion of the c-axis shows an anomalous S-shape, previously identified as a hallmark of the type-II superionic transition in fcc ice \cite{forestier_x-ray_2025}. 
This suggests a superionic transition in the hcp polymorph near $1700$ K, at the onset of the c-axis expansion in Fig. \ref{fig4}. The main difference with fcc ice is that the excess volume associated with H diffusion appears primarily accommodated by expansion along the $c$ direction, which may be related to anisotropic ionic conductivity. Interestingly, such an anisotropic proton hopping mechanism was discussed in the case of superionic \emph{pseudo}-hcp NH$_3$, associated with a clear elongation along the c-axis \cite{ninet_proton_2012}.
The temperature at which hcp ice enters a superionic state is similar to the case of fcc ice \cite{forestier_x-ray_2025}. 

The combined results from runs 1 and 2 (Fig. \ref{fig1}) suggest a revised high-pressure high-temperature phase diagram of H$_2$O. In run 1, the hcp phase appears as a transient state between fcc and bcc upon cooling, and our data do not support its thermodynamic stability in this regime. In run 2, within the $160-200$ GPa range at high temperatures, coexistence of hcp and fcc ice is observed with hcp dominating the diffraction signal (Fig. \ref{fig3}, bottom pattern).
In our previous study in Ref. \onlinecite{forestier_x-ray_2025}, fcc ice was observed as the dominant phase in this domain. However, in light of the present results, a diffraction peak previously marked with an asterisk (see Fig. 2 in Ref. \onlinecite{forestier_x-ray_2025}) and attributed to an unidentified reflection likely corresponds to the 101 reflection of hcp ice. This peak was observed above $130$ GPa in the data of Ref. \onlinecite{forestier_x-ray_2025}. This suggests that hcp ice was already present in our earlier experiment, although it was not interpreted as such at the time. A coherent picture therefore emerges: the fcc–hcp transition proceeds via a martensitic mechanism within the high-temperature superionic domain and extends over a broad pressure range, represented as a color gradient in Fig. \ref{fig1}. The relative phase proportions may vary between experiments and are likely influenced by the thermodynamic path, kinetic effects, and non-hydrostatic stresses, as documented for compressed Xe \cite{shimizu_pressure-induced_2008, cynn_martensitic_2001}. Preferential orientation, as discussed in Ref. \onlinecite{brugman_strength_2025}, may also account for differences in the observed phase proportion between experiments.
Recently, Andriambariarijaona \emph{et al.} \cite{andriambariarijaona_observation_2025} reported a mixed close-packed ice structure manifested as a broad diffraction signal around the fcc 111 reflection, from reverberating-shock experiments coupled with ultrafast XRD. This fcc-hcp coexistence observed at $150-180$ GPa and $2500$ K is consistent with the phase diagram disclosed in Fig. \ref{fig1}. 
Finally, at 219 GPa (Fig. \ref{fig3}, top pattern), the hcp peaks clearly dominate, suggesting that the relative stability of the hcp sublattice increases beyond 200 GPa and eventually surpasses that of fcc in the superionic state.
Similarly to the quenched fcc phase reported in Ref. \onlinecite{forestier_x-ray_2025}, hcp ice was recovered at ambient temperature in run 2 (Fig. S7 of the SM). Although simulations predict an enthalpy excess of $0.3$ eV at $0$ K relative to ice X \cite{cheng_phase_2021}, the experimentally observed density of the quenched hcp phase is higher than that calculated in Ref. \onlinecite{cheng_phase_2021}, which may enhance its stability, as previously suggested for the fcc phase \cite{forestier_x-ray_2025}.

The properties of superionic warm dense ice have been discussed in models explaining the non-dipolar, non-axisymmetric magnetic fields of Uranus and Neptune \cite{stanley_convective-region_2004,millot_experimental_2018,militzer_phase_2024,matusalem_plastic_2022}. The present observation that an hcp form of superionic ice may exist deeper within the inner shell of ice giants could therefore have important implications for the stratification and dynamics of their superionic mantles, as it may possess mechanical and electrical transport properties distinct from those of the commonly assumed fcc structure \cite{matusalem_plastic_2022}. 
Our findings should motivate further theoretical work on the physical properties of hcp ice---especially its mechanical plasticity and electrical conductivity. Finally, continued experimental efforts are needed to more comprehensively establish the pressure-temperature stability domains of hcp and fcc ice phases.

\begin{acknowledgments}
We acknowledge the European Synchrotron Radiation Facility (ESRF) for provision of synchrotron beamtime under proposals number HC5078, and HC5699. The authors acknowledge the Agence Nationale de la Recherche (ANR) for financial support under Grant No. ANR-21-CE30-0032-01 (LILI).
\end{acknowledgments}

\bibliographystyle{apsrev4-2}

%

\end{document}